# The using of bibliometric analysis to classify trends and future directions on "Smart Farm"

Paweena Suebsombut[1,2], Aicha Sekhari[2], Pradorn Sureepong[1], Pittawat Ueasangkomsate[3] and Abdelaziz Bouras[4]

[1]College of Arts Media and Technology, Chiang Mai University, Chiang Mai, Thailand
{Paweena.Suebsombut@univ-lyon2.fr}{dorn@camt.info}
[2]University Lumiere Lyon 2, Bron, France {aicha.sekhari@univ-lyon2.fr}
[3]Kasetsart University, Bangkok, Thailand {pittawat.uea@gmail.com}
[4]Qatar University {abdelaziz.bouras@qu.edu.qa}

*Abstract*—Climate change has affected the cultivation in all countries with extreme drought, flooding, higher temperature, and changes in the season thus leaving behind the uncontrolled production. Consequently, the smart farm has become part of the crucial trend that is needed for application in certain farm areas. The aims of smart farm are to control and to enhance food production and productivity, and to increase farmers' profits. The advantages in applying smart farm will improve the quality of production, supporting the farm workers, and better utilization of resources. This study aims to explore the research trends and identify research clusters on smart farm using bibliometric analysis that has supported farming to improve the quality of farm production. The bibliometric analysis is the method to explore the relationship of the articles from a co-citation network of the articles and then science mapping is used to identify clusters in the relationship. This study examines the selected research articles in the smart farm field. The area of research in smart farm is categorized into two clusters that are soil carbon emission from farming activity, food security and farm management by using a VOSviewer tool with keywords related to research articles on smart farm, agriculture, supply chain, knowledge management, traceability, and product lifecycle management from Web of Science (WOS) and Scopus online database. The major cluster of smart farm research is the soil carbon emission from farming activity which impacts on climate change that affects food production and productivity. The contribution is to identify the trends on smart farm to develop research in the future by means of bibliometric analysis.

## I. INTRODUCTION

Smart farm, defined by Alliance for Internet of Things Innovation (AIOTI), is data gathering, data processing, data analysis and automation technologies application on the overall value chain. This unique approach allows management and operation improvement of a farm with respect to real time and re-use of these data including animal, plant, and soil to improve food security which refers to the awareness and prevention of foodborne illnesses from food production to consumption, and chain optimization to deliver more productive and sustainable farm production based on more accurate and resource-efficient approach supported by Internet of Things (IoT) technologies [1]. From the farmer's point of view, smart farm should provide added value of productivity to farmers that are strongly related to cultivation and livestock farm in both large and small-scale farmers having new and more precise tools to produce more productivity, and also provide the benefits in terms of environmental issue [1]. Therefore, smart farm is the management optimization of inputs in a farm field according to veritable needs of crop that are composed of data-based technologies, including remote sensing, internet, and satellite-positioning systems like GPS, to manage crops and to reduce the use of water, fertilizers, and pesticides.

According to the prediction of the Food and Agriculture Organization (FAO), the world population will increase by about 1,500 million people in the next 40 years that emphasizes on the necessity and importance to increase the awareness on quality, quantity, and sufficiency of food for the population in the world [2]. Balancing between population growths and the food supply is the aim of food security. Food safety includes food loss which refers to obtaining an unusual reduction in quality such as wilting, contusion or lost before reaching the consumer. Also, food waste which refers to good quality of food and suitable for consumption, but it is discarded. Food loss normally occurs in the food value chain at the production stage, storage stage, processing and distribution stages, and is the unpremeditated result of infrastructure, technical limitation or farm processes in storage, packaging and marketing. Food waste generally occurs in the food value chain at the consumption and retail stages, and is the result of a conscious decision or negligence when throwing food away [1]. The transparency of food safety should become near the real-time which is quite challenging. Therefore, new chain cooperation and applications can lead to a more responsive and dynamic food production network. This can be accomplished by an attempt on making an appropriate distribution of food to all parts of the world.

But the food safety is at risk due to unpredictable environmental situation, such as climate change, which is a major problem for cultivation and livestock. Climate change can create unwanted risks to the environment. The consequences of experiencing heavy rainfalls, more extreme droughts, flooding, and the increasing in emissions of greenhouse gases resulting from human activities can abruptly change the seasonal life cycle events of plants and animal livestock [2]. From the Intergovernmental Panel in Climate Change (IPCC) report in 2010, the emissions of Global Greenhouse Gas by forestry, agriculture, and other land use sector is around 24% excluding $CO_2$ [3]; this is shown in figure1. Mostly greenhouse gas emissions from agricultural sector occur from agriculture in both cultivation and livestock, and deforestation to make way for cultivation of crops and livestock. Moreover, the shifting distribution of pests and diseases also has an impact on food production in the future.

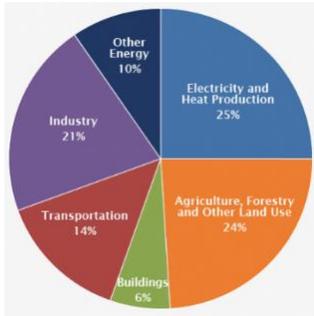

**Figure 1.** Global Greenhouse Gas Emissions [3]

From these impacts there is a negative result in the limited number of crops grown and starvation for the livestock which is raised. One way to counter the risk from devastating the operation of farming is to apply the concept of smart farm technology, which is used in farms to analyze and manage farms by following the farms' conditions. Smart farm technology is composed of a variety of different technological implementations and information, such as smart sensing and monitoring, smart control, could-based computing, smart analysis and planning that are applied in a variety of areas in the world, which has an effect of climate change for improving production. Therefore, a country does not have to be concern about the increase of food production which could further increase the emissions of global greenhouse gas from agricultural activities. Meanwhile, changing the way of production in farm is one solution for many food producers for adapting to the climate change. In developing countries, the gross domestic product (GDP) of agriculture sector accounts for 29% and provides jobs for 65% of their populations [4]. From a climate change viewpoint, smart farm is important and is adopted in every country to improve food production and productivity. In European nations more than 200 projects related to smart farm are supported by the Horizon 2020 idea which is a European funding initiative in agriculture and forestry conservation. In 2016-2017, the Horizon 2020 offers a budget of €560 million to boost farm productivity and sustainability [5]. The aims are to improve the capacity of food and farming systems, providing sufficient and healthy food, while safeguarding natural resources. The projects relate to applying the IoT technologies in smallholders and large farms for gathering data from farm, analyzing, and farm management. Australia is also one of the world leaders in agricultural innovation. Farmers in Australia adopt smart technologies that are supported by researchers, industry groups, and other stakeholders. This enthusiasm for change and innovation has helped Australian agriculture to retain its competitive edge over other producers. The exports in agriculture sector of Australia are more than $40 billion value of goods in each year and are essential to continuing economic strength of Australia. To assure Australia's ongoing competitiveness, the Government plan to grow agriculture by investing $4 billion to their farmers for helping farmers innovate an extension of the Rural R&D, thus providing agreements of export trade can help farmers to get benefit of the trade opportunities, and providing more encourage to communities can help farmers to build resilience of drought [6].

The USA spent a lot of money to improve farm production and productivity. For The Fiscal Year 2015, The United States Department of Agriculture (USDA) is the U.S. federal executive department responsible for executing and developing federal laws which relates to farming, agriculture, food, and forestry that spend around $140 billion budget for the agricultural sector [7]. Furthermore, the Department of Agriculture (USDA) has a budget $24.6 billion to invest in rural communities, rural youth, start-up projects for farmers and ranchers, agricultural research in nutrition assistance, food safety, vulnerable populations, and natural resources [8]. Farming is the major source of food and avocation for lots of poor families especially in Asia to contribute the economic output. One of third accounting of the Gross Domestic Product (GDP) of Myanmar, Lao, and Cambodia in 2010 is 39.9%, 30%, and 28% respectively. In Cambodia, over 60.3% of the labor force is provided, which is close to percent of the labor force in Vietnam (49.5%), and over percent of the labor force in Thailand, Indonesia, and Philippines which are 41.3%, 38.3%, and 33.6% respectively in 2010 [9]. From the effect of climate change, the Intergovernmental Panel in Climate Change (IPCC) finds that the agriculture sector of Indonesia, the Philippines, Thailand and Vietnam can be prospective to result 2.2% in the GDP of mean drop in 2100 [9]. Therefore, countries in Asia try the adoption of technology that is geared towards improving production and productivity: such an example is Japan which paid $46.5 billion in subvention to farmers and continued, whereas the state help of farmers in Japan remains a argumentative issue in 2009 [10]. The key goal in agricultural sector of African Union is to establish 25 million climate smart farmers by 2025 to achieve shared prosperity and improve the livelihoods of the African population [11]. The technologies are adopted to improve their productivity and sustainable production.

Furthermore, several countries also try to promote smart farm into an organic farm, especially in the development for sustainability farming that can reduce insecticides and pesticides usage which have a negative effect to the environment, consumers, and food safety. Almost 44 million hectares of agricultural land is operated in the form of an organic farm in 172 countries whereas 11 countries have more than 10 percent organic agricultural land [12]. The most organic agricultural land in 2014 is Australia with 17.2 million hectares. The main organic crops are cereals, vegetables, green podders, oilseeds, coffee, dried pulses, olives, grapes, nuts, and cacao [12].

This study reveals and explores the smart farm that has supported farming in the world including researches, projects, advantages, and limitation of smart farm. It consists of four further sections. The research trends on smart farm will be presented in section 2 which explains each cluster of smart farm by using biliometric analysis. Section 3 is a use case of smart farm which describes and gives an example of a use case of adoption smart farm into a real field. Finally, the conclusion and discussion are described in section 4.

## II. RESEARCHES TRENDS ON SMART FARM

This section presents the methodology for clustering research trends by using bibliometric analysis which can be used to analyze the scientific domain [13] following 3 steps by using the VOSviewer tool. The first step is conducting a data collection from an online database that are related to research articles on smart farm, agriculture, supply chain, knowledge management, traceability, and product life cycle management from the ISI WOS online database over the year 1991-2016 and Scopus online database over the year 1960-2016 comprising information for analysis about journals article, theses, and books. The next step, Co-citation unit analysis is selected to measure the relationship between cited documents for clustering research on smart farm. Finally, the bibliometric map is created to understand a research trends and to explain their relationship. The history of the number of published articles on smart farm issues from 1998 to 2016 (access on January to November 2016) is illustrated on Figure2. The publications in this research field increased from 1 article in 2007 to 14 articles in 2009 then suddenly decreased in 2010 and increased again from 18 articles in 2012 to 75 articles in 2015. There were 75 articles in 2015 and 54 articles from January to November 2016 published in this field. The total number of published articles in this field from 2008 to 2016 is 287 articles with a 93.53% increasing rate of article published. From 294 selected articles, the frequencies of citation list are illustrated in Table1 that shows the number of citations for the top ten cited articles. The articles are often referenced in research on soil Carbon Emissions from farming activity and Food Security and Farm management.

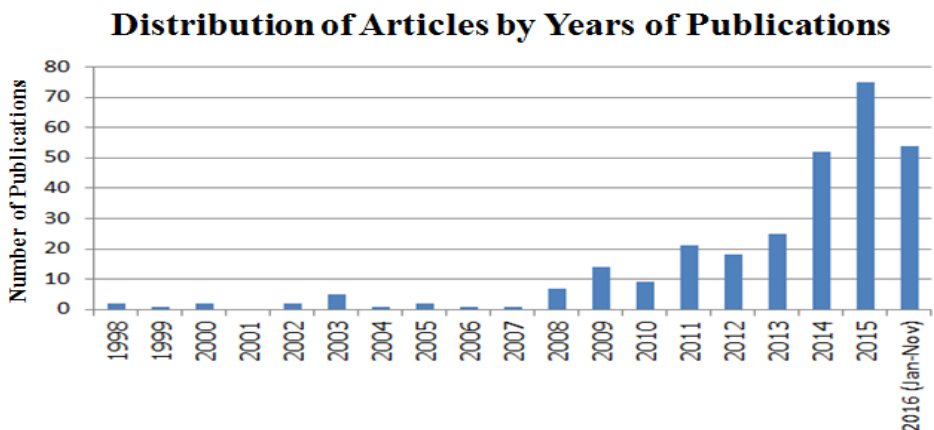

**Figure2.** Distribution of articles by Year Publications

**Table1.** Top ten of frequencies of citation

| Topics | Number of citations |
|---|---|
| Lal, Rattan. "Soil carbon sequestration impacts on global climate change and food security." science 304.5677 (2004): 1623-1627. | 3237 |
| Doran, John W., and Timothy B. Parkin. "Defining and assessing soil quality." Defining soil quality for a sustainable environment defining soil qua (1994): 1-21. | 2414 |
| Batjes, Niels H. "Total carbon and nitrogen in the soils of the world." European journal of soil science 47.2 (1996): 151-163. | 2370 |
| West, Tristram O., and Wilfred M. Post. "Soil organic carbon sequestration rates by tillage and crop rotation." Soil Science Society of America Journal 66.6 (2002): 1930-1946. | 1496 |
| Tilman, David, et al. "Global food demand and the sustainable intensification of agriculture." Proceedings of the National Academy of Sciences 108.50 (2011): 20260-20264. | 1328 |
| West, Tristram O., and Gregg Marland. "A synthesis of carbon sequestration, carbon emissions, and net carbon flux in agriculture: comparing tillage practices in the United States." Agriculture, Ecosystems & Environment 91.1 (2002): 217-232. | 880 |
| Reeves, D. W. "The role of soil organic matter in maintaining soil quality in continuous cropping systems." Soil and Tillage Research 43.1 (1997): 131-167. | 864 |
| Paustian, K., et al. "Management options for reducing CO2 emissions from agricultural soils." Biogeochemistry 48.1 (2000): 147-163. | 647 |
| Gartner, Tracy B., and Zoe G. Cardon. "Decomposition dynamics in mixed-species leaf litter." Oikos 104.2 (2004): 230-246. | 644 |
| Flato, Gregory, et al. "Evaluation of Climate Models. In: Climate Change 2013: The Physical Science Basis. Contribution of Working Group I to the Fifth Assessment Report of the Intergovernmental Panel on Climate Change." Climate Change 2013 5 (2013): 741-866. | 543 |

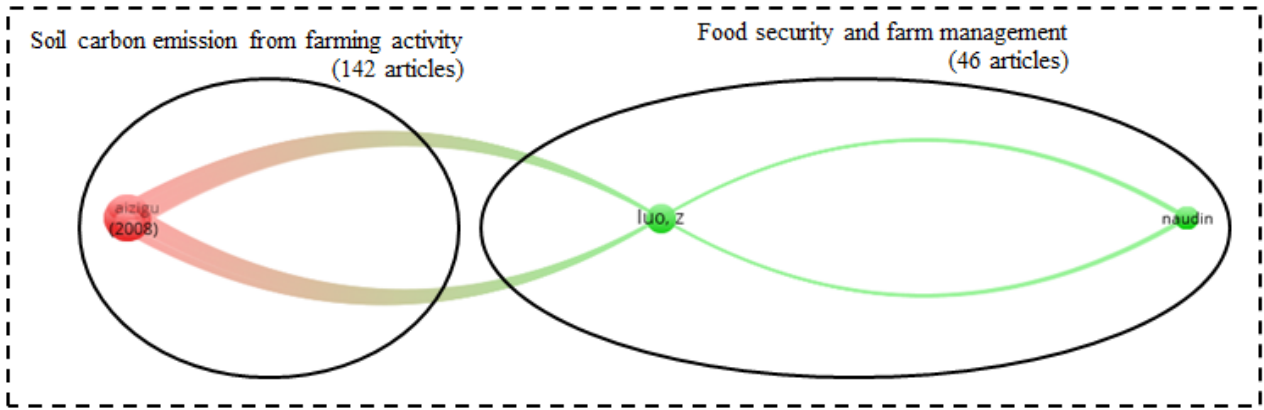

**Figure3.** Two clusters for research on smart farm

Figure3 illustrated the co-citation map is created by the VOSviewer tool. The total of 188 articles are found from co-citation network that can be classified in two clusters; Soil Carbon emissions from farming activities, and food security and farm management. Each cluster is demonstrated in a relationship between the cited documents via number of frequencies that is shown in Table1. Therefore, the research field of soil carbon emission from farming activities is the major cluster that can explore the research trends in a smart farm.

*A. Soil Carbon Emissions from farming activity*

In this cluster, the researchers focus on the emissions of soil carbon which have occurred from farming activities. The emission of soil carbon and modern farming are the major drivers of climate change and is accelerating the rate of biodiversity loss that have already exceeded the boundaries of the Earth [14] that affects food security.

The carbon sink capacity of soil degradation which is a soil quality in terms of the capacity of the soil to produce goods or services [15] and farming activities of the world is 50 to 66% of the carbon loss in history which is 42 to 78 gigatons (Gt) of carbon [16]. The Emissions of the soil carbon from farming activities are relative with cultivation practice and crop type which are generated from three sources which are production and fertilizers application, pesticides, and irrigation, using machine for cultivating the land, and the soil organic carbon (SOC) that is oxidized from soil infestation [17]. The world soil carbon pool of 2500 Gt includes about 950 Gt of soil inorganic carbon (SIC) and 1550 Gt of soil organic carbon (SOC). Conversion of natural to farming ecosystems also causes a reduction of the SOC pool by 75% or more in cultivated soils of the tropics regions and 60% in soils of temperate regions [16].

The Carbon emission from the machinery used for fertilizers, pesticides, and irrigation depends on amounts of fertilizers, pesticides, and irrigation of each crop that varies among crop rotations, crop types, and tillage practices [17].

The rotation of crop can influence the soil Carbon stocks by changing the quality and quantity of organic substance input. All operation for tillage affects to Carbon emission involving mechanical soil disturbance for seedbed preparation depending on various factors comprising properties of soil, tractor size, depth of tillage and implementation used [18]. The irrigation is significant to achieve higher yields in semiarid and arid regions. The irrigated cropland is 17% leading to 40% of total production on a global scale. The Carbon emission in irrigation is a very precise practice which is 23% of the on-farm energy used in being estimated for crop production. For example, energy requirement of on-farm pumping in the USA for pumping water depends on various factors composing of the total dynamic head which is based on system pressure, water lift, the rate of water flow, pipe friction, and the efficiency of pumping system [19].

Therefore, decreasing losses from enhancing the use in efficiency of all these inputs can help to reduce Carbon emission indications. The management options that contribute to $CO_2$ emissions from farming activities are reducing or eliminating soil tillage (no-till) and increasing cropping intensification and plant production efficiency. No-till promotes the formation of insubordinate soil organic matter (SOM) fractions within stabilized micro-aggregate and macro-aggregate structures, and increases aggregate stability [20].

*B. Food Security and Farm management*

This cluster includes two sub-clusters that are food security and farm management. Food security is a major priority at national and household level for the poor and political stability welfare. The demands of consumer are presently the main drivers to encourage food industries to produce healthier and safe food products for the highest possible quality specifications. Governments representing their developed country have adopted several strategies comprising of government markets intervention, endeavors to enhance production, and public distribution of food and preservation of national food security stocks [21]. To develop agriculture sector, there have commitments of $20 billion over three years.

Furthermore, the funding approaching $30 billion in the period 2010-2012 resulted from the commitments of the Copenhagen Accord, and by 2020 the goal of funding and additional $100 billion annually for helping developing countries react to climate change comprising adaption and mitigation for getting more productivity and sustainable [22].

To secure the farm productivity, the farm management is important. Currently, the management tasks in farm are moving to a new model, requiring more empathy on the interaction with the environmental effect, and quality of documentation and growing conditions. Generally, the change of conditions for the management tasks in farm has enforced the monitoring system activities and information systems to secure allowance with the limitations and standards of particular production guidelines, management standards as prerequisites for subsidies, and provisions for environmental approbation [23]. Most farmers have handled with management load by trying to manipulate manual information's mass for making the correct decisions. So that, computers and the internet usage can help to enhance and mitigate the task of manipulating and internal information processing as well as achieving external information [23]. Several methods and technologies are used to improve farm management that helps farmers to make a decision, manage farms, and improve farm production and productivity. In North China, over 50% nation's total area is dryland regions where low fertility soil, adverse weather, water resource and topography conditions are the limitation of farming development. The conservation tillage in dryland regions is developed by reduction tillage practice that helps to increase water use efficiency and crop yields with up to 35% [24]. The European countries have an infrastructure planned for future internet by the European Future Internet Initiative (EFII) by merging Public–Private Partnership (FIPPP) with a purpose of essentially developing the implementation and realization of Future Internet services and establishing a European scale markets by integrating communications functionality for smart infrastructures by 2015 [25]. Farm Management Information System (FMIS) which is a system to collect, process, store, and disseminate is necessary information to manage the functions of farm operations, collects farm profile and data related to product traceability from consumers that applies advanced Future Internet characteristics. This work attempts to capacitate a farmer to pace into a new verity, where they become a veritable. The Future Internet services are pioneered optimized methods providing to manage all tasks in a farm for entering the global markets [25]. This cluster includes a research done on food security and its method for farm management that can help farmers manage their farms more efficiently.

III. USE CASES OF SMART FARM

There are numerous cases used on smart farm that every country tries to adopt for improving their farming practice in cultivation and livestock. Countries in Asia are launching national strategies to promote the farm automation with data analytics, robotics, and sensor technology. These tools can make a major difference to crop yields, quality of crop, and profits of rural farmers. Australia uses robotic into dairy operations which is a project in Camden, South-West of Australia that 90 cows can be milked an hour. The duration and volume of its last milking are recorded and transmitted from a dongle that is put around the cow neck. So, farmers can access all data and monitor yield and production records via their iPads [26]. Japan tries to apply robots to automate crop farms and pick fruits. The temperature, humidity, carbon dioxide levels, light sources, and sterilizing water are controlled by automated system. The ripe strawberries and tomatoes can be recognized by robot fruit picker machines based on the color. The robot will calculate the distance between arms and fruit for harvesting without bruising the fruit, and harvest faster than farmers [26]. In Thailand, researchers in Mahidol University adopted drones to monitor crop conditions, ground robots to measure the soil nutrients, and weather forecast system to report farmers about changing of weather that can help farmers to reduce wastage in agriculture, to reduce the farming cost, and to decrease the effect on the environment [26]. The Fuse Technologies platform is a global collaboration between DuPont Pioneer and AGCO Corporation including European countries, the United States, Brazil, and Canada which is a global leader in the design, manufacture and distribution of agricultural equipment. The wireless data transfer technology is a solution for leading markets in agriculture. The data interface and farm management information are required for the system to make the whole-farm decision which is designed to help farmers enhance their productivity and profit. The task file management system allows farmers securely and appropriately into transferring task files between station and farm machines via wireless transferring [27]. A part in the North West of Italy applies a wireless sensor network controlling irrigation and enhances fertilization methods on corn crops that are composed of wind vane, anemometer, and pluviometer sensors. The conditions of micro-local weather, air temperature, direction and intensity of wind, rain level, wetness of leaf, moisture and temperature of soil at 0.5 meters depth can be monitored [28]. The Smart Agriculture project for Organic Farms in UK was deployed in 9 farms located across the UK in which gainful data for predicting measurements and events are recorded via wireless sensor network and GPRS communication that can be helped agronomists and farmers saving money and time [29].

Consequently, various smart farm technologies and management are adopted in farming to control and improve production and productivity that can help farmers to reduce time, increase farmers' profits, reduce carbon emission. However, the smart farm still has a limitation for farming depending on the restriction of each country such as the absence of physical communication infrastructures especially in rural areas, knowledge technology of farmers, cost of devices, marketing, and supply chain management. All countries try to reduce the limitation for the farmers and infrastructure.

## IV. CONCLUSION AND DISCUSSION

This paper revealed the current trends on smart farm including cultivation and livestock in both the research project and on the cases applied in the world. The research articles which are discussed in the paper are collected from Scopus and ISI WOS online databases. The first result comprises the distribution of articles based on the yearly publications. The number of publication on smart farm issue has grown evidently from 1998 to 2016 (access on January to November 2016). The relationship between the articles is measured via a co-citation unit by using a VOSviewer tool. The results of this mapping show the relationship between the cited articles with two clusters that are soil carbon emission from farming activity, and food security and farm management. The major trends on smart farm is related to soil carbon emission from farming activity that has an impact on the climate change, which also includes applying smart farm to mitigates the effect from climate change and to increase the quality of soil and farm productivity following the food security concept. The limitation of this paper is presenting the research trends on smart farm based on model of co-citation mapping. For future research, the statistical analysis will be improved to confirm this relationship.

## ACKNOWLEDGEMENT

The authors would like to acknowledge the support of University Lumiere Lyon 2, France, College of Arts, Media and Technology, Chiang Mai University, Thailand, Kasetsart University, Bangkok Thailand, and Qatar University, Qatar. We also would like to acknowledge all of our colleagues who worked together and provided encouragement to the authors.